%% file: beta.tex
\documentclass[12pt]{iopart}
\usepackage{iopams}  
\usepackage{epsfig,rotating}

\input mydef.tex

\begin{document}
\title{Physics Reach of the Beta Beam}

\author{Mauro Mezzetto}

\address{Istituto Nazionale Fisica Nucleare, Sezione di Padova, Italy.
\vskip0.5cm Invited talk at the Nufact02 Workshop, Imperial College of Science, Technology and Medicine, London,
July 2002.}

\begin{abstract} 
Beta Beams are designed to produce pure (anti)electron neutrino beams and could
be an elegant and powerful option for the search of leptonic CP violating
processes.
In this paper will be quantified the physics reach of a CERN based Beta Beam
and of a Super Beam -- Beta Beam combination.
The CP phase $\delta$ sensitivity results to be comparable to a Neutrino
Factory for $\sin^2{\theta_{13}}$ values greater than $10^{-4}$.
\end{abstract}

%Uncomment for PACS numbers title message
%\pacs{00.00, 20.00, 42.10}

% Uncomment for Submitted to journal title message
%\submitto{\JPA}

% Comment out if separate title page not required
\maketitle

\section{Introduction}
The Beta Beam concept has already been presented in \cite{Piero}, more
details on the acceleration scheme can be found in \cite{Lindroos}.
The possibility to produce a pure \nubare\  or \nue\  beam from 
 \He\  or \Ne\  ion beams opens a unique opportunity in the long term search for
leptonic CP violation.

This paper is focussed to a well defined scenario, namely a Beta Beam produced
ad CERN and fired at a gigantic water \v{C}erenkov detector, \`a la
UNO (440 kton fiducial volume) \cite{UNO}, located under the Frejus at a baseline of 130 km.

The intrinsic capability to detect \thetaot and the
CP phase $\delta$  with a pure Beta Beam will be studied
 together with the possibility to detect
T, CP and CPT violating effects combining Beta Beam with the CERN
SPL SuperBeam \cite{SPL-SB} fired to the same detector.

\section{Signal and backgrounds} 
The signal in a Beta Beam looking for \nuenumu oscillations would be the
appearance of \numu\  charged-current events, mainly via quasi-elastic
interactions. These events are selected by requiring
        a single-ring event,
        the track identified as a muon using the standard 
Super-Kamiokande identification algorithms, and
        the detection of the muon decay into an electron.
The backgrounds and signal efficiency have been studied in a
full simulation, using the NUANCE code~\cite{casper}, reconstructing
events in a Super-Kamiokande-like detector.

The Beta Beam is intrinsically free from contamination by any different
flavour of neutrino. However, background can be generated by inefficiencies
in particle identification, such as single-pion
production in neutral-current (NC) \nue\ (\nubare) interactions, electrons
(positrons) mis-identified as muons, or by external sources such as
atmospheric neutrino interactions.

%
%In a water \v{C}erenkov detector pions below 1~GeV can be distinguished
%from muons only by requiring the decay electron signature to be associated
%to an identified muon. 
%
%
Electrons (positrons) produced by \nue\  (\nubare) can be mis-identified
as muons, therefore giving a fake signal.
Standard algorithms for particle identification in water \v{C}erenkov
detectors are quite effective to suppress such backgrounds. Furthermore,
the signal of the decay electron in muon tracks can be used to reinforce
the muon identification.
%A full simulation of these backgrounds predicts an electron rejection
%of the order of xxx \% for a mean efficiency in background detection
%of xxxx.
%

Atmospheric neutrino interactions are estimated to be $\sim$~50/kton/yr in
the energy range of interest for the experiment, a number of interactions
that far exceeds the oscillation signal. The atmospheric neutrino
background has to be reduced mainly by timing of the parent ion bunches.
 For a decay ring of
6.9~km and a bunch length of 10~ns, which seems feasible \cite{Lindroos}, 
 a rejection factor of $2\cdot 10^4$ can be achieved. The directionality of the
(anti)neutrinos can be used to suppress further the atmospheric neutrino
background by a factor $\sim 4$. With these rejection factors, the
atmospheric neutrino background can be reduced to the order of 1 event/440
kton/yr. Moreover, out-of-spill neutrino interactions can be used to
normalize this background to the 1\% accuracy level.

\section{Beam optimization}

The Lorentz boost factor $\gamma$ of the ion accelerator can be tuned to
optimize the sensitivity of the experiment. 
Optimization is performed assuming $\delta m^2_{23} =
2.5\cdot10^{-3}$~eV$^2$, a baseline of
130~km  and a $^6$He beam. In principle,
baselines in the range 100-250~km are possible considering that
at present SPS can accelerate \He\  ions up to $\gamma=150$. 

The number of quasi-elastic events in the far detector scales roughly
as $\gamma^{3}$ (
this factor comes from the
beam focusing, which is $\propto \gamma^2$, and the cross section, 
which is $\propto \gamma$) and so
the number of quasi-elastic events at a given L/E is proportional
to $\gamma$. 
\begin{figure}[th]
  \epsfig{file=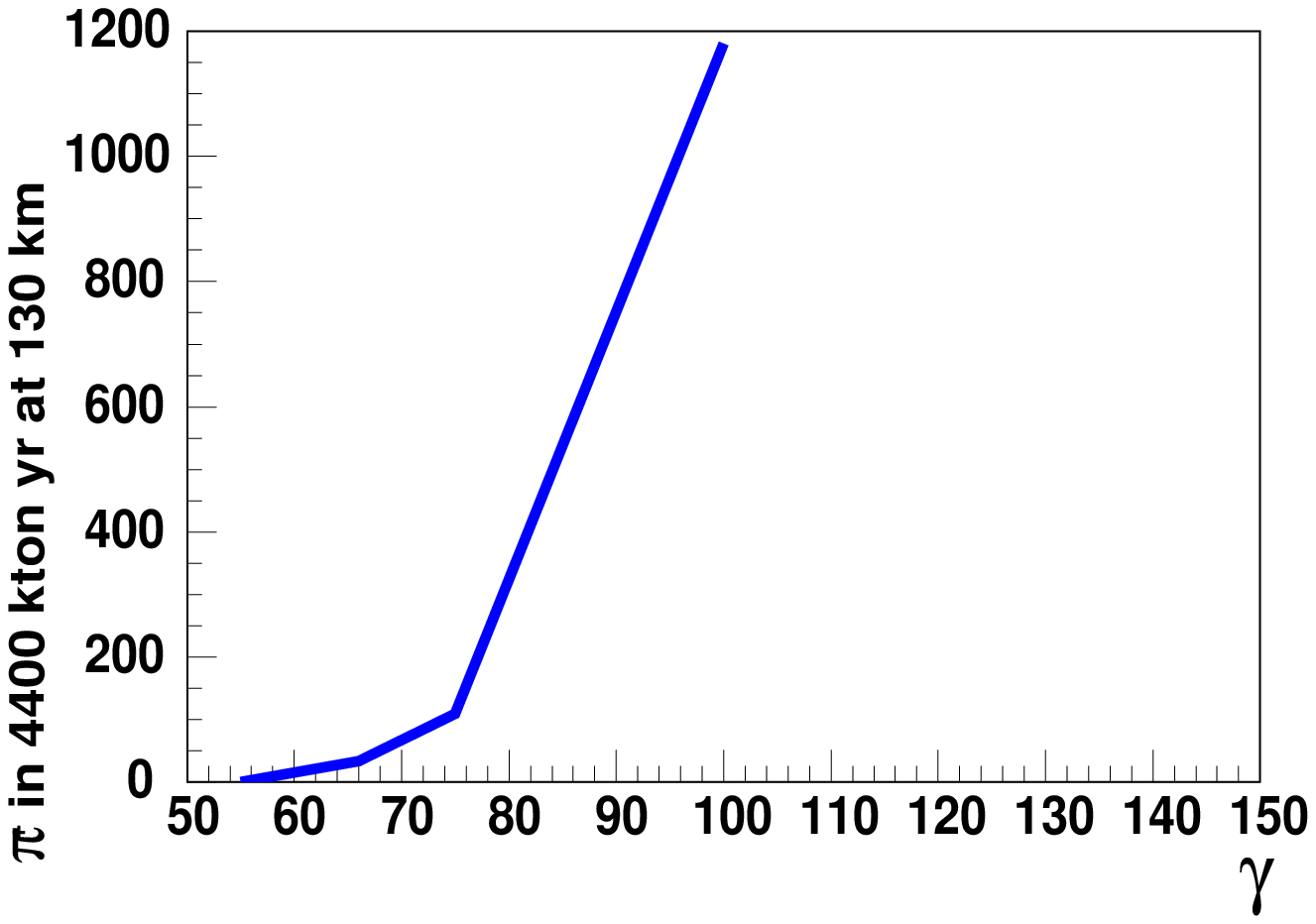,width=0.33\textwidth}
  \epsfig{file=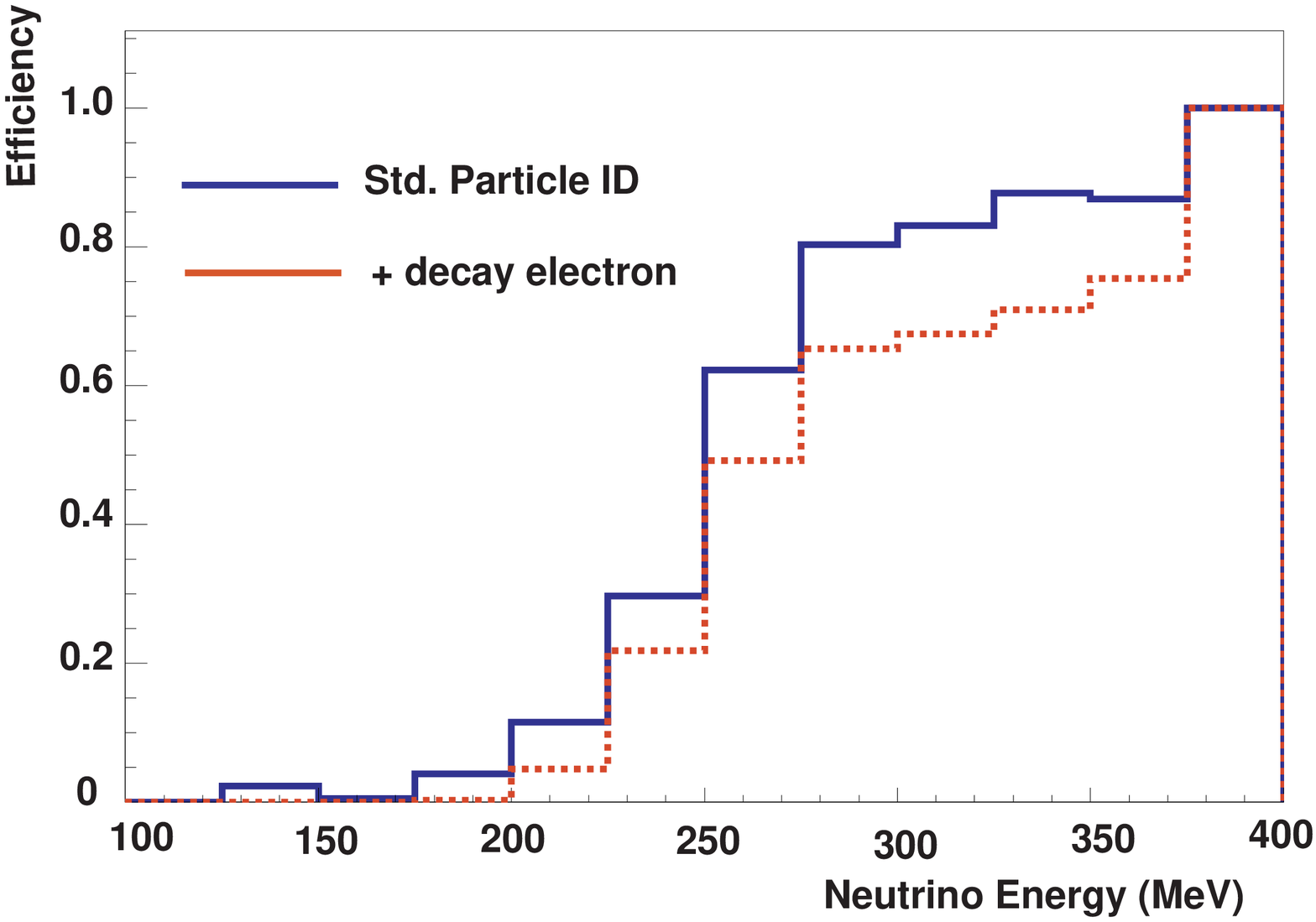,width=0.33\textwidth}
  \epsfig{file=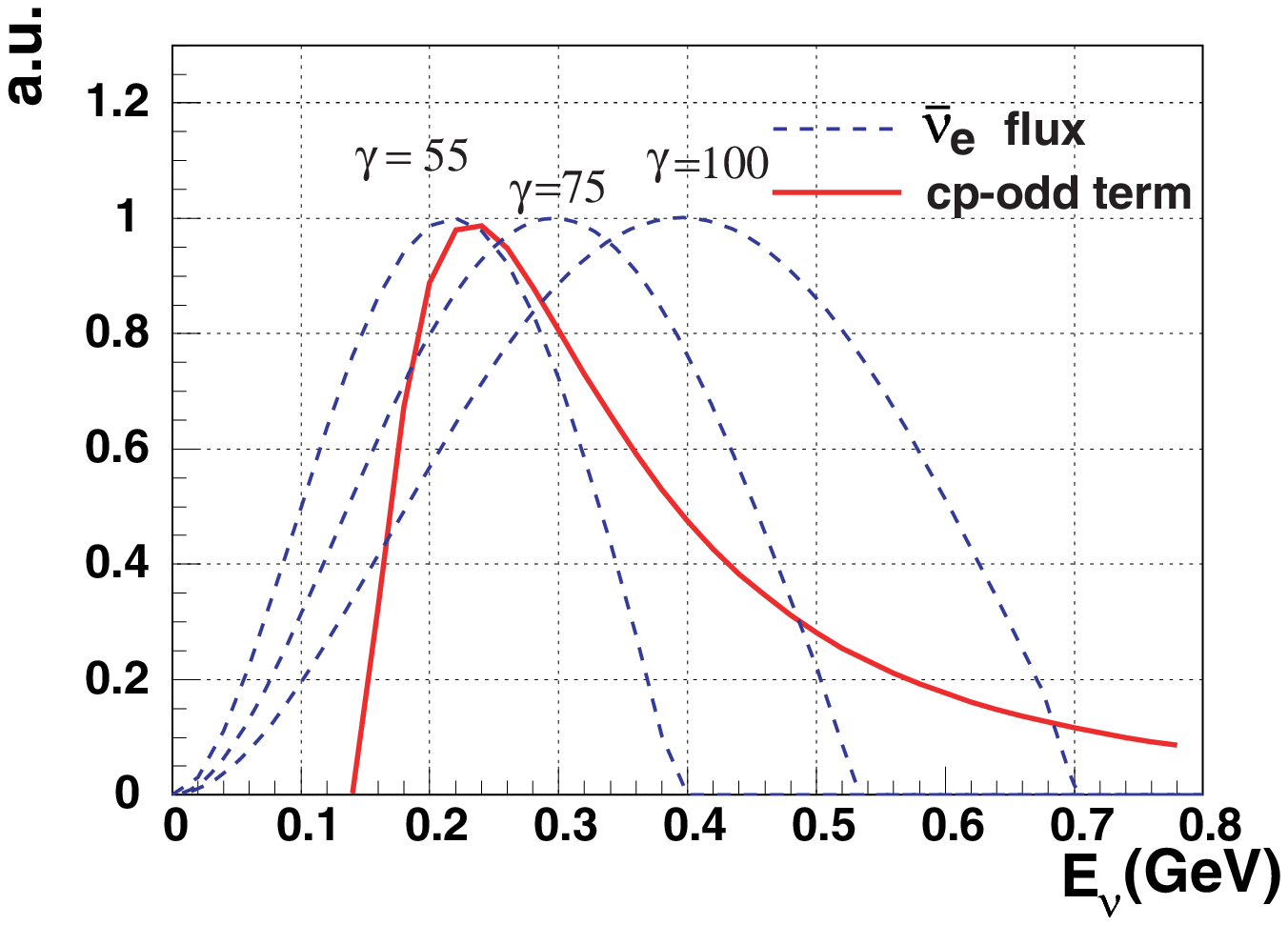,width=0.33\textwidth}
\caption{Left panel: backgrounds from single-pion production for a 
4400~kt-y
exposure, as function of $\gamma$. 
Middle panel: signal efficiency as function of the neutrino energy 
computed with and without the detection of the decay electron.
Right panel:
 energy spectra for different values of $\gamma$ compared with 
the CP-odd term of $p(\nubare\rightarrow\nubarmu)$.}
\label{fig:beta:eff}
\end{figure}

On the other hand
the number of background events from NC pion production increases
very rapidly
with $\gamma$, as shown in Fig.~\ref{fig:beta:eff}(left).
The threshold for this process is about 400 MeV and for
$\gamma <55$ no pions are created above the \v{C}erenkov
threshold.

A third factor is
the signal efficiency as function of energy, 
Fig.~\ref{fig:beta:eff}(middle), which severely disfavours
neutrino energies below 300~MeV. 
The signal efficiency decreases slightly above 0.7~MeV where
the fraction of quasi-elastic, single ring events decreases.

Finally
the true signal of the experiment, 
\numu\  produced by the CP-odd term in \pnuenumu,
should be matched by the neutrino energy spectrum.
It is evident from Fig.~\ref{fig:beta:eff}(right) that this is
no longer true 
when $\gamma \geq 100$.

Based on these considerations, a $\gamma$ value of 75 seems to approach
the optimal value for CP sensitivity.
For $\gamma=75$ we will assume a flux of $2.9 \cdot 10^{18}$ \He\ 
decays/year and $3.6\cdot10^{17}$ \Ne\  decays/year, as
discussed in \cite{Piero}.
Table~\ref{tab:beta:rates} reports signal and background rates for a 
4400~kt-y exposure to  $^6$He and $^{18}$Ne beams.

\section{Systematic errors}

The cross-section estimates for signal and background production at
energies below 1 GeV are quite uncertain, the systematic errors being of
the order of 20 and 30\% respectively. 
On the other
hand, a Beta Beam is ideal for measuring these cross sections, provided
that
a close detector of 1~kton at least is placed at the distance of 
about 1~km from the decay tunnel.

The energy and the flux of the neutrino beam is completely defined
by the acceleration complex and
the near/far residual error is extremely reduced, because in a Beta Beam
the divergence
of the beam is completely defined by the decay properties of the parent
ions.
The $\gamma$ factor of the accelerated ions can be varied, in particular
a scan can be initiated below the background production threshold,
allowing a precise measurement of the cross sections for resonant processes.
It is estimated that a residual systematic error of 2\% will be
the final precision with which both the signal and the backgrounds
can be evaluated.

\thetaot and $\delta$ sensitivities are computed taking into account
        a 10\% error on the solar $\delta m^2$ and $\sin^2{2\theta}$,
as expected from the Kamland experiment after 3 years of data 
taking~\cite{KamLAND} and
  a 2\% error on the atmospheric $\delta m^2$ and $\sin^2{2\theta}$,
as expected from the JHF neutrino experiment~\cite{jhf}.
Only the diagonal contributions of these errors are considered in the
following. 

Correlations between \thetaot and $\delta$ are fully accounted for,
and indeed they are negligible in this configuration,
while the sign of $\delta m^2_{13}$ and the
$\theta_{23}/(\pi/2-\theta_{23})$ ambiguities are not considered.

\section{Sensitivity to CP violation}

A search for leptonic CP violation can be performed running the Beta Beam
with $^{18}$Ne and $^6$He, and fitting the number of muon-like events to
the \pnuenumu probability.
 The fit can provide simultaneous determinations
of $\theta_{13}$ and the CP phase $\delta$.
Given the relative interaction rates for quasi-elastic events, a sharing
of 3 years of $^6$He and 7 years of $^{18}$Ne has been considered.

\begin{figure}[ht]
\centerline{\epsfig{file=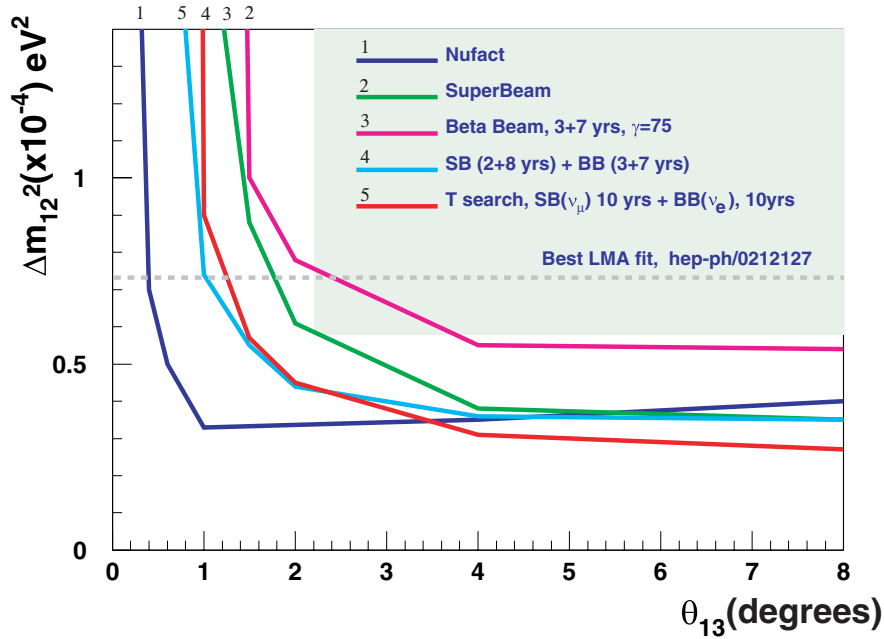,width=0.80\textwidth}  }
\caption{CP sensitivity of the Beta Beam, of the SPL-SuperBeam, 
and of their combination, see text. Two different time sharing
are considered: 3+7 years of \nubare, \nue\  Beta Beam combined
with 2+8 years of \numu, \nubarmu\  Super Beam and 10 years of \nue\  Beta Beam
with 10 years of \numu\  Super Beam. Sensitivities are 
 compared with a 50 GeV Neutrino Factory 
producing       $2\cdot10^{20} \mu$ decays/straight section/year, 
and two 40 kton detectors at 3000 and 7000 km \cite{golden}.
The shaded region corresponds to the allowed LMA solution and the
\thetaot sensitivity of JHF.}
\label{fig:beta:sens}
\end{figure}

The results of this analysis are summarized in Table~\ref{tab:beta:rates},
for an arbitrary choice of the mixing matrix parameters.
\begin{fulltable}{\label{tab:beta:rates}
%\begin{table}
%\caption{\label{tab:beta:rates}
 Event rates for a 4400~kt-y exposure. The signals are
computed for $\thetaot=3^\circ$, $\dmot=0.65\cdot10^{-4}eV^2$, 
$\sin^2{2\theta_{12}}=0.8$,
$\delta=90^\circ$.}
\lineup
\begin{tabular}{@{}lrrrr}
\br
10 years        & \multicolumn{2}{c}{Beta Beam} & \multicolumn{2}{c}{SuperBeam}\\
 		& $^6He$($\gamma=75$)  &   $^{18}Ne$($\gamma=75$) & $\pi^+$ focus & $\pi^-$ focus    \\
\hline
CC events (no osc, no cut) & 40783   &  18583 & 183488 & 29150\\
Total oscillated           &  32  &   66 & 565 & 85\\
CP-Odd oscillated          & -47   &  24 & -390 & 61\\
Beam background            &  0  &  0  & 703 & 126\\
Detector backgrounds       &  60  &  10 & 182 & 62\\
\br
\end{tabular}
\end{fulltable}
%Fig.~\ref{} shows, for comparison with the SPL-SuperBeam, how
%would be solved 6 possible combinations of \thetaot and $\delta$.
Since the sensitivity to CP violation is heavily dependent on the true 
value of \dmot and \thetaot, we prefer to express the CP sensitivity for a
fixed value of $\delta$ in the \dmot, \thetaot parameter space.
The CP sensitivity to separate
$\delta=90^\circ$ from $\delta=0^\circ$ at the 99\%CL as a function
of \dmot and \thetaot, following the convention of~\cite{golden},
is plotted in Fig.~\ref{fig:beta:sens}.

The \He\  Beta Beam  sensitivity to \thetaot,
computed for $\delta=0$ and solar SMA solution, is 
$\thetaot \geq 1.2^\circ \quad (90\% \rm{CL})$ in a 2200 kton/years
exposure.

\section{Synergy between the SPL-SuperBeam and the Beta Beam}

The Beta Beam needs the SPL as injector, but consumes at most $\sim 3\%$
of the SPL protons. The fact that the average neutrino  energies of
both the SuperBeam and the Beta Beam are below 0.5 GeV, with the beta
beam being tunable, offers the fascinating possibility of exposing the
same detector to two neutrino beams at the same time.

The SPL-SuperBeam is a \numu\  (\nubarmu) beam, while 
the Beta Beam is a
\nue\  (\nubare) beam and so the combination of the two offers
the possibility of CP, T and CPT searches at the same time:
  \begin{itemize}
        \item
    Searches for CP violation, running the SuperBeam with \numu\  and 
\nubarmu, and the Beta Beam with $^6$He (\nubare) and $^{18}$Ne (\nue).
        \item
    Searches for T violation, combining neutrinos from the SuperBeam 
(\numunue) and from the Beta Beam using $^{18}$Ne ($\nue \rightarrow 
\numu$), or antineutrinos from the SuperBeam ($\nubarmu \rightarrow 
\nubare$) and from the Beta Beam using $^6$He ($\nubare \rightarrow 
\nubarmu$).
        \item
    Searches for CPT violation, comparing $P(\nu_\mu \rightarrow \nu_e)$ to
$P(\nubare \rightarrow \nubarmu)$ and $P(\nubarmu \rightarrow \nubare)$
to $P(\nue \rightarrow \numu)$.
\end{itemize}
It is evident that the combination of the two beams would not result
merely in an increase in the statistics of the experiment, but would offer
clear advantages in the reduction of systematic errors, and would offer 
the redundancy needed to establish firmly any CP-violating effect within
reach of the experiment.

%Given the neutrino fluxes and cross-sections, for the same level of systematic errors, 
%    the most powerful combination of beams would be however a single T search
%    with neutrinos (SuperBeam \numu\  with BetaBeam \nue).

It should also be stressed that  the Super+Beta beams offer the
only known possibility of measuring CPT violation combining signals 
from the same detector, a crucial issue for the control of systematic
errors.

Fig.~\ref{fig:beta:sens} summarizes the CP sensitivity of different
combinations of Super Beam and Beta Beam compared with the Neutrino
Factory sensitivity. The combined sensitivity 
shown in Fig.~\ref{fig:beta:sens} is competitive with
that of a Neutrino Factory, offering a truly complementary approach to the
search of leptonic CP violation.
In particular
in the parameter space defined by the solar LMA solution and
the JHF phase I sensitivity on \thetaot ($\thetaot \geq 2.2^\circ$) the
Beta+Super Beam sensitivity is very similar to the Neutrino Factory 
sensitivity. We consider solar LMA plus a \thetaot value within the
reach of JHF the only possible trigger for a Leptonic CP search.

Since the boundary condition of this study is quite arbitrary, namely a
440 kton detector and Beta Beam fluxes extrapolated from the present
knowledge of the production rate \He\  and \Ne\  ions (cfr. ref.~\cite{Piero}),
it is of interest to study the Super+Beta beam capabilities with a bigger
detector or with higher \Ne\  fluxes. Statistics and \nue\  fluxes are
indeed the two major bottlenecks of Beta Beam.
\begin{figure}[th]
\centerline{\epsfig{file=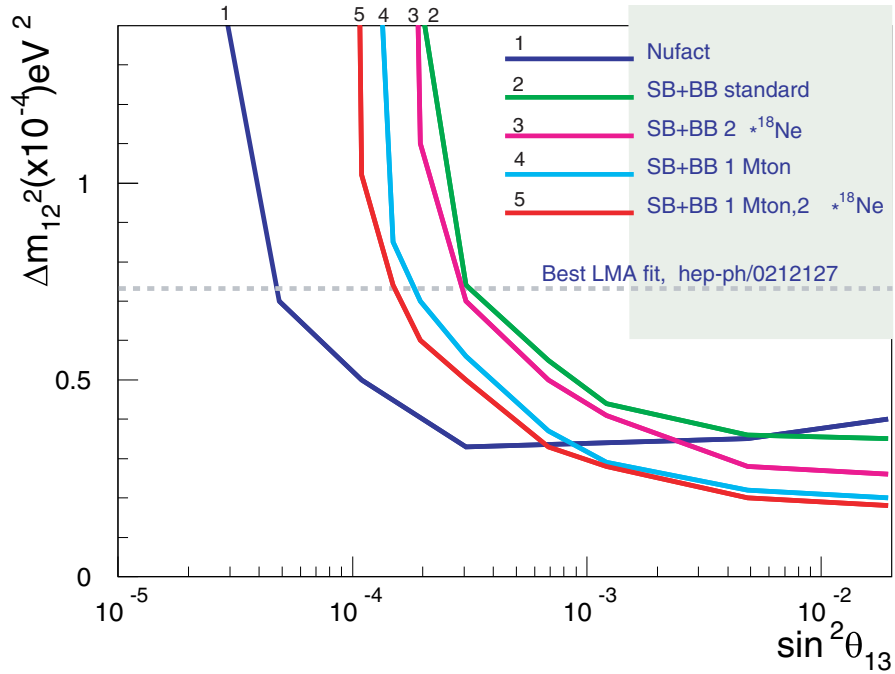,width=0.80\textwidth} }
\caption{CP sensitivity of the combination of the Beta Beam with
 the SPL-SuperBeam, for different choices of the detector size
and \Ne\  fluxes, see text. 
Time sharing is 3+7 years of \nubare, \nue\  Beta Beam combined
with 2+8 years of \numu, \nubarmu\  Super Beam. 
 Sensitivities are 
 compared with a 50 GeV Neutrino Factory 
producing       $2\cdot10^{20} \mu$ decays/straight section/year, 
and two 40 kton detectors at 3000 and 7000 km \cite{golden}.
The shaded region corresponds to the allowed LMA solution and the
$\sin^2{\thetaot}$ sensitivity of JHF.}
\label{fig:CP:improvements}
\end{figure}
Improvements on CP sensitivity with a 1 Mton detector, \`a la HyperK
\cite{jhf} and with a \Ne\  flux increased by a factor two, are
shown on Fig.~\ref{fig:CP:improvements}.
Another possible improvement could be the exploitation of the event
spectral distribution, not studied here, by using the algorithms 
studied in ref.~\cite{Campanelli}.

\end{document}

%% file: mydef.tex
\def\nue{\ensuremath{\nu_{e}}}
\def\nubare{\ensuremath{\overline{\nu}_{e}}}

\def\numu{\ensuremath{\nu_{\mu}}}
\def\nubarmu{\ensuremath{\overline{\nu}_{\mu}}}

\newcommand{\pnuenumu}{\ensuremath{p(\nue \rightarrow \numu)\,}}
\newcommand{\nuenumu}{\ensuremath{\nue \rightarrow \numu\,}}

\newcommand{\dmot}{\ensuremath{\delta m^2_{12}\,}}

\newcommand{\He}{\ensuremath{^6{\mathrm{He}\,}}}
\newcommand{\Ne}{\ensuremath{^{18}{\mathrm{Ne}\,}}}

\newcommand{\thetaot}{\ensuremath{\theta_{13}}\,}
\newcommand{\numunue}{\ensuremath{\nu_\mu \rightarrow \nu_e}}